\documentstyle{aipproc}
\def\be{\begin{equation}}
\def\ee{\end{equation}}
\def\bea{\begin{eqnarray}}
\def\eea{\end{eqnarray}}
\begin{document}
\title{Gauge Field Theories on a $\perp$ lattice}
\author{Matthias Burkardt}
\address{New Mexico State University\\
Las Cruces, NM 88003, U.S.A.}

\maketitle

\begin{abstract}
In these notes, the transverse ($\perp$)
lattice approach is presented as a means to control the $k^+\rightarrow 0$
divergences in light-front QCD. Technical difficulties of both the canonical 
compact formulation as well as the non-compact formulation of the $\perp$ 
lattice motivate the color-dielectric formulation,
where the link fields are linearized.
\end{abstract}

\section*{Introduction}
The main subject of these notes are difficulties associated with the
formulation of gauge field theories on a transverse ($\perp$) lattice
using light-front (LF) quantization. 
Because of these difficulties, the reader may wonder about the advantages
of this approach --- particularly given the successes of Euclidean
lattice gauge theory (LGT). 
The primary motivations for formulating QCD in this framework
is that LF quantization is the most physical approach towards a microscopic
description of the parton distributions measured in deep-inelastic scattering
as well as many other hard processes. \footnote{This and other motivations
are discussed in more detail in Ref. \cite{mbadv} and in references therein.}
It is important to emphasize this fact in this brief introduction since it 
explains why LF quantization of QCD and the $\perp$ lattice should be
investigated as a possible alternative to Euclidean and Hamiltonian LGT 
formulations --- despite the difficulties that will be discussed in the remainder 
of these notes.

\subsection*{Why LF gauge?}
Although the choice of quantization hyperplane and the choice of gauge are
in principle independent issues, the so-called LF gauge ($A^+=0$) turns out
to be highly preferable for the canonical formulation of LFQCD. The main
reason is that in the kinetic energy term for ${\vec A}_\perp$
(from $-\frac{1}{4}F^{\mu \nu}F_{\mu \nu}$)
\be
{\cal L}_{kin, A_\perp} =
D_+ {\vec A}_\perp D_- {\vec A}_\perp
= \left( \partial_+-igA_+\right){\vec A}_\perp
\left( \partial_--igA_-\right){\vec A}_\perp,
\ee
the term multiplying the `time' derivative of ${\vec A}_\perp$
(i.e. $\partial_+{\vec A}_\perp$) contains also $A_-=A^+$. Therefore, the
canonical momenta
\be\Pi = \frac{\partial {\cal L}}{\partial (\partial_+A_\perp)}=
\left( \partial_--igA_-\right) A_\perp,
\label{eq:Pi}
\ee which are the LF
analog to $\Pi = \frac{\partial {\cal L}}{\partial (\partial_0A_\perp)}$
in equal time quantization, are ``simple'' (i.e. linear in the fields)
if and only if $A_-=A^+=0$. Therefore, in order to avoid having to deal with a 
system
that has to satisfy nonlinear constraints 
\footnote{Eq. (\ref{eq:Pi}) is a constraint equation since it 
involves no time-derivative.}
one normally selects $A^+=0$ gauge before quantizing in LF coordinates.

However, this choice of gauge is not entirely free of problems.
To illustrate this fact, let us start from the Euler-Lagrange equation for $A^-$ in
QED \footnote{Since the problem that we are going to discuss occurs already in
QED, we will discuss it there because of the simpler algebra.}
\be
-\partial_-^2 A^- = g J^+ + \partial_-{\vec \partial}_\perp {\vec A}_\perp
\equiv g \tilde{J}^+ \label{eq:constr}
\ee
(the LF analog to the Poisson equation), which is also a constraint
equation. It is  convenient to eliminate $A^-$, using the
solution to Eq.(\ref{eq:constr}), i.e.
$
A^-(x^-,{\vec x}_\perp) = -\frac{g}{2}\int_{-\infty}^\infty
dy^- \left|x^--y^-\right| \tilde{J}^+(y^-,{\vec x}_\perp), 
$
yielding an instantaneous interaction term 
\be
V^{inst} = -\frac{g^2}{4} \int_{-\infty}^\infty \!dx^- 
\int_{-\infty}^\infty \!dy^- \int \!d^2x_\perp
\tilde{J}^+\!(x^-,{\vec x}_\perp )
\left| x^--y^-\right| \tilde{J}^+\!(y^-,{\vec x}_\perp ).
\ee
This linearly rising interaction in the LF Hamiltonian
causes IR divergences, unless
\be
\int_{-\infty}^\infty dx^-\tilde{J}^+(x^-,{\vec x}_\perp) = 0
\quad \quad \quad \quad\quad \quad
\forall {\vec x}_\perp .
\label{eq:gauss}
\ee
The origin of this
problem lies in the fact that setting $A^+=0$ does not completely fix the gauge
freedom. An $x^-$ independent gauge transformation
$
A^\mu \longrightarrow 
{\cal U}^\dagger A^\mu {\cal U} -\frac{i}{g} {\cal U}^\dagger \partial^\mu 
{\cal U}
,\mbox{with} \quad {\cal U}={\cal U}({\vec x}_\perp),
$
leaves $A^+=0$ unchanged and Eq. (\ref{eq:gauss}) is just the Gau\ss'  law 
constraint associated with this residual gauge symmetry\footnote{Fixing the 
remaining gauge freedom requires dealing
with explicit zero-mode degrees of freedom and it is still
not completely understood how to do this!}.
As long as ${\vec x}_\perp$ is a continuous variable, Eq. (\ref{eq:gauss}) 
implies an infinite
number of constraint on the states ($\infty$ number of ${\vec x}_\perp$!),
which is again difficult to deal with. This is one of the motivations
for discretizing the $\perp$ space direction in the context of LF 
quantization.

\section*{The transverse lattice}
The basic idea behind the $\perp$ lattice \cite{bardeen}
is to work in two continuous
($x^0$ and $x^3$ or $x^+$ and $x^-$) space time directions and two
discrete [${\vec x_\perp} \equiv (x_1,x_2)$] space directions, i.e. space-time
consists of a 2-dimensional array of 2-dimensional sheets. 
\footnote{In the closely related Hamiltonian LGT
space-time consists of a 3-dim. array of 1-dim. lines.} 
The motivation
for working with such a `hybrid' formulation is that the discretized 
$\perp$ directions provide a the possibility to introduce a gauge invariant
cutoff, while the continuous longitudinal directions allow to maintain
manifest longitudinal
boost invariance (which is one of the advantages of the LF formulation).

The natural way to introduce gauge fields within this framework seems to be to
work with compact link-fields $U_\perp\in SU(N_C)$ in the discretized
$\perp$ directions (as is done in conventional LGT)
and with non-compact gauge fields $A^\pm$ in the continuous longitudinal 
directions. It should be emphasized that both the $U_\perp$'s as well as
the $A^\pm$ (which are 
defined on the links and sites of the $\perp$ lattice respectively), are 
functions of two discrete and two continuous 
variables, i.e. one can think of the $\perp$ lattice action as consisting of 
many 1+1 dim. gauge theories coupled together.

The trouble with this formulation is the nonlinear $U_\perp\in SU(N_C)$ 
constraint on the
link fields. The reason that this constraint is more difficult to handle
on the $\perp$ lattice than in Euclidean or Hamiltonian LGT
is due to the fact that the $U_\perp$s are still two dimensional fields
(and not just variables, as in Euclidean LGT, or quantum mechanical
rotors, as in Hamiltonian LGT). Despite several attempts in this direction
\cite{paul}, nobody has been able to construct a Fock space basis out
of these ``nonlinear $\sigma$ model'' degrees of freedom which still
allows one to evaluate matrix elements of the LF Hamiltonian.

Two possibilities to avoid the problems associated
with the $SU(N)$-constraint have been pursued: The first is to work with
non-compact gauge fields also in the $\perp$ direction and the other
is to keep compact fields, but to relax the $SU(N)$ constraint and
linearize the degrees of freedom.

\subsection*{Non-compact formulation of the $\perp$ lattice}
Again, we illustrate the main difficulties in the context of QED.
In order to satisfy the $U(1)$ constraint, one starts with the ansatz 
$U_\perp= \exp \left(ieA_\perp\right)$, yielding
\be
P^- \sim 
-\frac{1}{4}\sum_{{\vec n}_\perp} \int \!\!dx^- \int \!\!dy^-
\tilde{j}^+(x^-,{\vec n}_\perp)
\left| x^--y^-\right| 
\tilde{j}^+(y^-,{\vec n}_\perp) + P^-_{plaq},
\ee
where $P^-_{plaq}$ is the $\perp$ plaquette interaction  ($xy$ orientation) and 
\be
\tilde{j}^+(x^-,{\vec n}_\perp) = j_q^+(x^-,{\vec n}_\perp)
+ \frac{1}{e} \Delta_\perp \partial_- A_\perp ,
\label{eq:9}
\ee
where $\Delta_\perp$ is the discrete approximation to the
$\perp$ Laplace operator and $j_q^+$ is the portion of the current
carried by the fermions.

As long as one restricts oneself to gauge fields with {\em local}
fluctuations only (as in the Fock expansion!) one finds 
$A_\perp (+\infty) = A_\perp (-\infty)$ and thus
\be 
\int_{-\infty}^\infty dx^- \tilde{j}^+(x^-,{\vec n}_\perp) = 
\int_{-\infty}^\infty dx^- j_q^+(x^-,{\vec n}_\perp).
\ee
Together with Gau\ss ' law this implies that 
$\int_{-\infty}^\infty dx^- j_q^+(x^-,{\vec n}_\perp)=0$, 
i.e. charges must add up to 
zero at each site! Transversely separated charges are only allowed
in the presence of ``soliton-like'' gauge fields with
$A_\perp(x^-=+\infty)-A_\perp(x^-=-\infty)= ke$, where $k$ is an integer!
The physics behind this result becomes clear by noting that
$\exp \left(ieA_\perp\right)$ acts like such a soliton operator: from the
canonical commutation relations
$\left[A_\perp(x^-), A_\perp(y^-)\right]=\frac{i}{2} \varepsilon (x^--y^-)$,
together with $\left[e^A,B\right]=\left[A,B\right]e^A$ if 
$\left[A,B\right]$ is a c-number,
one finds
\be
\left[A_\perp(x^-=+\infty)-A_\perp(x^-=-\infty)
, \exp \left(ieA_\perp(y^-)\right)\right]
= - e \exp \left(ieA_\perp(y^-)\right) .
\ee
Therefore the Gau\ss ' law constraint is satisfied if transversely
separated charges are separated by a string of exponentials ---
just as one would have expected from gauge invariance ---
and the good news is that the infrared divergences cancel if states
are constructed in a gauge invariant way. 

It is instructive to examine in detail how the $k^+\rightarrow 0$
divergences cancel in $QED_{2+1}$.
In 2+1 dimensions, there is only one $\perp$ direction and
therefore purely $\perp$ plaquette terms are absent. As a result,
the whole dynamics of $A_\perp$ is described by its coupling
to $A^-$ and pure gauge, coupled to external sources, becomes
exactly solvable in the non-compact formulation.
 
The rest frame energy of an external source $j^+$
in $QED_{2+1}$ on a $\perp$ lattice is given by
$
H_{RF} = v^+ P^- + H_{recoil},
$
where $v^+$ is the velocity of the source and
\be
P^-= \frac{e^2a}{2}\sum_n\int_{-\infty}^{\infty}dq^+
\tilde{j}_n(q^+)\frac{1}{{q^+}^2}\tilde{j}_n(-q^+).
\label{eq:pjj1}
\ee
is the instantaneous interaction arising from eliminating $A^-$ and
\be
H_{recoil} = \frac{1}{2v^+} \sum_n \int_0^\infty dk^+
a_n^\dagger (k^+) a_n(k^+) k^+
\ee
is a recoil term which appears in the LF description of fixed
sources \cite{mbzako}.
The effective current $\tilde{j}^+$ receives contributions from
both $A_\perp$ and the external current $j^+$ [see also Eq.(\ref{eq:9})]. 
In momentum space, one finds
for the current on the $n^{th}$ site
\be
\tilde{j}^+_n(q^+) = e j^+_n(q^+) + \frac{iq^+}{a}
\left[A_{\perp, n}(q^+) - A_{\perp, n-1} (q^+)\right],
\ee
where we define the $n^{th}$ site to be the one between 
the $(n-1)^{th}$ and the $n^{th}$ link.
It is instructive to decompose the instantaneous interaction into terms
quadratic in $A_\perp$ and $j^+$ respectively and a mixed term, i.e.
$P^-=P^-_{AA}+P^-_{jj}+P^-_{JA}$, where
\bea
P^-_{AA} &=&\sum_n\!\int_0^\infty \!\!\!\frac{\,dk^+}{2a^2k^+}
\!\!\!\left[a_n^\dagger(k^+) - a_{n-1}^\dagger(k^+)\right]\!\!\!
\left[a_n(k^+) - a_{n-1}(k^+)\right]  \nonumber\\
&=& \frac{4}{a^2}\!
\int_0^\infty \!\frac{dk^+}{2k^+}
\!\int_{-\pi/a}^{\pi/a} \!\!\!dk_\perp
\sin^2\!\left(\frac{ak_\perp}{2}\right)
a^\dagger(k^+,k_\perp)a(k^+,k_\perp).
\label{eq:qedfree}
\eea
where 
$
a_n(q^+) = \int_{-\pi/a}^{\pi/a} \frac{dq_\perp}{\sqrt{2\pi}}
a(q^+,q_\perp)\exp \left( iq_\perp a n \right)\,
$. Furthermore
\be
P^-_{jj}=
\frac{e^2a}{2}\sum_n\int_{-\infty}^{\infty}dq^+
j_n(q^+)\frac{1}{{q^+}^2}j_n(-q^+)
\label{eq:pjj2}
\ee
is the self-energy of the source, and the coupling of the
source to $A_{\perp}$ reads
\bea
P^-_{jA} &=&  -ie\sum_n\int_{-\infty}^{\infty}\frac{dq^+}{q^+}j_n(q^+)
 \left[\tilde{A}_n(q^+)-\tilde{A}_{n-1}(q^+)\right]
\label{eq:pja}
\\
&=& \frac{ie}{\sqrt{2a}}
\int_{0}^{\infty}\!\!\!\!\frac{dq^+}{{q^+}^{3/2}} 
\left\{j^+_n(q^+)
\left[ a_n(q^+) - a_{n-1}(q^+)\right]
- j^+_n(-q^+)
\left[ a_n^\dagger(q^+) - a_{n-1}^\dagger(q^+)\right]\right\}. \nonumber
\eea
In order to calculate the self-energy of an external charge-distribution
$j_q(q^+)$ to order $e^2$, one needs to add the instantaneous self-interaction
(\ref{eq:pjj1}) [which is of ${\cal O}(e^2)$ already] in first order to the
contribution from the coupling to $A_\perp$ (\ref{eq:pja}) [which is only
${\cal O}(e)$] treated in $2^{nd}$ order perturbation theory. The latter 
yields 
\begin{eqnarray}
\delta E^{(2)} &=& - \frac{{v^+}^2e^2}{a^2} \int_0^\infty \frac{dq^+}{{q^+}^3}
\int_{-\pi/a}^{\pi/a} \!\!dq_\perp
\frac{ 2 \sin^2 \left( \frac{a q_\perp}{2} \right) \tilde{j}(q)\tilde{j}(-q)}
     { \frac{q^+}{2v^+} + \frac{2v^+}{a^2q^+} \sin^2 \left( \frac{a q_\perp}{2} 
\right)},
\label{eq:deltaE2}
\end{eqnarray}
where we used the shorthand notation $j(q)\equiv j(q^+,q_\perp)$ and where
$
j_n(q^+) = \int_{-\pi/a}^{\pi/a} \frac{dq_\perp}{\sqrt{2\pi}}
j(q^+,q_\perp)\exp \left( iq_\perp a n \right)\,
$.
In general, $\delta E^{(2)}$ behaves for $q^+\rightarrow 0$ like
\be
\delta E^{(2)}_{div} =-\frac{v^+e^2}{2}\int_{-\infty}^\infty 
\frac{dq^+}{{q^+}^2}
\int_{-\pi/a}^{\pi/a} \!\!dq_\perp
j(q)j(-q) = -\frac{v^+e^2a}{2}\int_{-\infty}^\infty 
\frac{dq^+}{{q^+}^2}
\sum_n
j_n(0)j_n(0),
\label{eq:deltaE2div}
\ee
which diverges unless the net charge $j_n(0)$ on each (!) site is zero.
However, a similar divergence (with opposite sign) arises from the
instantaneous interaction, as can be directly read off from
Eq. (\ref{eq:pjj2}). The sum of the two terms is IR finite
as long as the total (i.e. sum of charge on all sites)
charge is zero\footnote{This is not surprising since $QED_{2+1}$ confines.}
\bea
\label{eq:de0}
\delta E &\equiv& v^+P^-_{jj} + \delta E^{(2)}
= \frac{e^2a}{2} \int_{-\infty}^\infty \frac{dq^+}{v^+}
\int_{-\pi/a}^{\pi/a} dq_\perp
\frac{\left(\frac{a}{2}\right)^2
j(q)j(-q)}{ \left(\frac{a{q^+}}{2{v^+}}\right)^2 + \sin^2 \left( \frac{a 
q_\perp}{2} \right)}.
\eea
In particular, for the case of two (oppositely charged)
point charges one finds in the limit $a\rightarrow 0$
the logarithmic interaction energy, characteristic for
an Abelian gauge theory in $2+1$ dimensions
\be
\delta E =\frac{e^2}{2\pi} \log \sqrt{R_\perp^2 + R_L^2} + \mbox{const.}\, .
\ee
At first this result (i.e. perfect cancellation of the IR
singularity in perturbation theory) seems to contradict 
the general discussion of the non-compact
formulation of gauge theories above, where it is shown that transversely
separated charges need to be connected by a gauge string in order to cancel
the IR divergences.

However, this apparent contradiction is resolved by
the simple observation the $\perp$ lattice Hamiltonian for non-compact 
$QED_{2+1}$
coupled to external sources is, technically speaking, just a bunch of
coupled shifted harmonic oscillators (the Hamiltonian is Gaussian!).
This has two important consequences:
First of all, for a shifted harmonic oscillators,
the exact ground state energy is obtained already in $2^{nd}$ order 
perturbation theory. Therefore, the calculated
ground state energy (\ref{eq:de0}) is the exact one.
Secondly, in the language of Fock space operators, the eigenstate
of a shifted harmonic oscillator are coherent states, i.e. exponentials
of raising operators.

Keeping these facts in mind, everything fits together:
The coherent states, which {\em are} the eigenstates of the
the $\perp$ lattice in the presence of the external source, 
accomplish the same effect as the exponentials of link fields, namely
the act as soliton like operators which are necessary to cancel the
small $q^+$ divergence of the instantaneous self-interaction.
Furthermore, the fact that we are dealing only with shifted harmonic 
oscillators in $QED_{2+1}$ also guarantees that we observed this
cancellation already in perturbation theory --- even though we did 
not construct these coherent states explicitly.

However, the bad news is that it is difficult to
construct a Fock space basis containing exponentials of the gauge fields.
\footnote{Even in $QED_{2+1}$, the exact solution contains exponentials of
$a^\dagger$ and thus leads to non-normalizable states.}
Furthermore, when one cannot solve the problem exactly, it is very difficult
to achieve this cancellation of IR singularities --- unless one works
with a basis of gauge invariant states, which is very hard in the
non-compact formulation.
As usual, in QCD, the situation is worse
because the gauge fields themselves carry
color-charge. Furthermore, if one tries to maintain gauge invariance then the
ansatz $U_\perp=\exp \left(igA_\perp\right)$ leads to
$\frac{1}{g^2}\partial_\mu \exp(igA_\perp) \partial^\mu \exp(-igA_\perp)
= \partial_\mu A_\perp \partial^\mu A_\perp +\mbox{``higher orders''}$ and
these higher order terms make it again very difficult to quantize the 
theory.
For these reasons, the non-compact formulation of the
$\perp$ lattice has been abandoned.
\subsection*{Color-Dielectric Formulation of the $\perp$ Lattice}
Naively, one might think that one possibility to introduce 
$U_\perp \in SU(N)$ fields would be to work with linearized general complex
matrix fields and to add a potential $V_{eff}(U_\perp)$ that has a minimum
for $U_\perp \in SU(N)$ (e.g. ``Mexican hat'' type potential). However, in the
Fock expansion (an essential ingredient in the LF Hamiltonian formulation),
the fields are usually expanded around the origin (or some other fixed value).
Expanding around the origin makes no sense for a Mexican hat shaped potential
since the origin corresponds to the false vacuum. On the other hand, expanding
around any point along the minimum breaks the manifest global gauge symmetry.
Therefore, this idea of adding a (Lagrange multiplier) effective self interaction
to enforce the constraint must be abandoned as well.

A solution out of this dilemma is to work with blocked (smeared, averaged)
degrees of freedom ${\cal M}$, which are obtained from the original
$U_\perp \in SU(N)$ by averaging $U_\perp$s or strings of $U_\perp$s
over some finite volume, e.g. by defining
${\cal M}=\sum_{av}U$ \cite{dalley}.
The advantage of this procedure is that the smeared ${\cal M}$s are no longer
subject to the strict $SU(N)$ constraint. 

The blocked theory is still equivalent to the original theory, provided
the action for the ${\cal M}$s contains an effective interaction {\it defined} 
by integrating out the $U_\perp$s
\be
\exp \left[-V_{eff}({\cal M})\right]
= \int {\cal D}U_\perp \delta \left( {\cal M}-\sum_{av}U\right)
\exp \left[-S_{canonical}(U_\perp)\right] .
\label{eq:veff}
\ee
The catch in the whole procedure is that  $V_{eff}({\cal M})$ as defined through
Eq. (\ref{eq:veff}) can be infinitely complicated and for its exact determination
one would have to perform a path integral. However, for approximate calculations,
one can always make an ansatz for $V_{eff}({\cal M})$ and there exist various
options to determine the parameters appearing in this ansatz.
Note that a direct use of Eq. (\ref{eq:veff})
to calculate $V_{eff}({\cal M})$ within the LF framework does not seem to be
possible: in order to evaluate the r.h.s. of Eq. (\ref{eq:veff}) one needs
to work with link-fields $U_\perp \in SU(N)$ and the difficulties in doing this
were the main motivation to introduce the color-dielectric formulation in the first 
place. In the Euclidean, calculating $V_{eff}$ seems to be more
straightforward, but for using it on the $\perp$ lattice, the
issue arises of translating $V_{eff}$ from the Euclidean to the LF.
\footnote{See Ref. \cite{mb:sg} for a discussion of this issue in the context of
scalar field theories.}
An alternative procedure, based on covariance requirements, appears to be very 
promising. Due to lack of space, the reader is referred to Refs. \cite{d2,dalley},
where the procedure has been discussed in detail.\\
{\bf Acknowledgements:} I would like to thank the organizers for the invitation
and DFG for financial assistance. It is a pleasure to thank P. Griffin, 
B. vande Sande and S. Dalley for many enlightening discussions.

\end{document}